\newcommand{\cjaa}{\textit{Chin. J.Astron.Astrophysics}}

\documentclass{iaus}

\usepackage{graphicx}

\title[Open clusters] %% give here short title %%
{Open clusters: their kinematics and metellicities}

\author[L. Chen, J.L. Hou, J.L.Zhao \& R. de Grijs ]   %% give here short author list %%
{Li Chen$^1$, Jinliang Hou$^1$, Junliang Zhao$^1$
 \and Richard de Grijs$^2$}

\affiliation{$^1$Shanghai Astronomical Observatory, Chinese
Academy of Sciences, 80 Nandan Road, Shanghai 200030, China
\\ email: {\tt chenli@shao.ac.cn, houjl@shao.ac.cn, jlzhao@shao.ac.cn} \\[\affilskip]
$^2$Department of Physics and Astronomy, University of Sheffield,
Hicks Building, Hounsfield Road, Sheffield S3 7RH, UK; and
National Astronomical Observatories, Chinese Academy of Sciences,
20A Datun Road, Chaoyang District, Beijing 100012, China
\\email: {\tt r.degrijs@sheffield.ac.uk}}

\pubyear{2008}
\volume{248}  %% insert here IAU Symposium No.
\jname{A Giant Step:from Milli- to
Micro-arcsecond Astrometry}
\editors{A.C. Editor, B.D. Editor \&
C.E. Editor, eds.}
\begin{document}

\maketitle

\begin{abstract}
We review our work on Galactic open clusters in recent years, and
introduce our proposed large program for the LOCS (LAMOST Open
Cluster Survey). First, based on the most complete open clusters
sample with metallicity, age and distance data as well as
kinematic information, some preliminary statistical analysis
regarding the spatial and metallicity distributions is presented.
In particular, a radial abundance gradient of - 0.058$\pm$ 0.006
dex kpc$^{-1}$ was derived, and by dividing clusters into age
groups we show that the disk abundance gradient was steeper in the
past. Secondly, proper motions, membership probabilities, and
velocity dispersions of stars in the regions of two very young
open clusters are derived. Both clusters show clear evidence of
mass segregation, which provides support for the ``primordial''
mass segregation scenarios. Based on the great advantages of the
forthcoming LAMOST facility, we have proposed a detailed open
cluster survey with LAMOST (the LOCS). The aim, feasibility, and
the present development of the LOCS are briefly summarized.

\end{abstract}

\firstsection % if your document starts with a section,
\section{The open cluster system and the observational database}

Open clusters (OCs) are considered excellent laboratories for
studies of stellar evolution. Studies in the research area dealing
with OCs showed a rapid growth in 1990's and this area continues
to develop vigorously. There may be many reasons for this recent
growth of OC studies. Many new techniques are greatly beneficial
for OC observations, including, for example, the application of
wide-field, high quality CCD cameras and, more recently,
multi-object spectrographs. In addition, OC studies play a very
important and unique role in determining the structure and
evolution of the Galactic disk.

OCs have long been used to trace the structure and evolution of the
Galactic disk. From an observational point of view, there are some
important advantages of using OCs as opposed to field stars. In
general, in terms of OCs, we deal with groups of stars of nearly the
same age, a similar composition and at a similar distance. We can
observe OCs to large distances, and the distances to most of the OC
sample have already been derived. In particular, OCs will have
relatively stable orbital motion, which can be used as a better
tracer of the Galactic disk structure. OCs also have a wide range of
ages, so that -- combined with their spatial distribution and
kinematic information -- we can study dynamical evolution effects,
and very young OCs are ideal objects for studies of the stellar
initial mass function. Furthermore, when combined with abundance
data, we can investigate the chemical evolution history of the
Galactic disk (\cite[Carraro et al. 1998]{Carraro_etal98},
\cite[Chen et al. 2003]{Chen_etal03}).

At present, the total number of detected clusters and associations
is around 1700 (\cite[Dias et al. 2002]{Dias_etal02}), about $60\%$
of which have distance and age information and for about half of
which proper motions are available. Less than one-fourth of the
sample have both proper motion and radial velocity parameters, and
only a small portion (about $8\%$) of OCs have well-determined mean
abundance values. We have compiled an updated OC catalog (Chen et
al., in prep.), for which data have been extracted from various
sources -- mainly from the Dias' catalog (\cite[Dias et al.
2002]{Dias_etal02}), Kharchenko's data(\cite[Kharchenko et al.
2005a]{Kharchenko_etal05a}, \cite[Kharchenko et al.
2005b]{Kharchenko_etal05b}) and the WEBDA database (\cite[Mermilliod
\& Paunzen 2003]{MermilliodPaunzen03}). Based on this large sample,
we can embark on statistical studies of the Galactic OC system.

\section{Spatial and kinematic properties of Galactic open clusters}

Using our updated database for 993 OCs with distance and age data,
we plotted the cluster positions on an $(X, Y)$ coordinate system,
with the zero point in $X$ at the Galactic Center (where $R_\odot =
8.0$ kpc), as shown in Fig. 1. The solid arc in Fig.1 represents the
solar circle about the Galactic Center. One can see from this figure
that -- apparently -- there are very few OCs at galactocentric
distances of less than 5kpc. Furthermore, while young clusters (with
age of less than 0.8 Gyr; see \cite[Phelps et al.
1994]{Phelps_etal94}) are distributed quite uniformly around the
Sun, most older OCs are distributed in the outer part of the disk.
These apparent distributions are partially due to the much higher
extinction in the direction of the Galactic Center and the
deficiency of older clusters in the inner part of the disk has been
also attributed to the preferential destruction of these clusters by
giant molecular clouds, which are primarily found in the inner
Galaxy.

Regarding the spatial distribution perpendicular to the Galactic
plane, most OCs represent the typical thin-disk population, with a
small scale height of about 66 pc. However, the subsample of old
OCs, most of which are found in the outer disk, has a much greater
scale height of 221 pc. These scale heights are in excellent
agreement with the earlier results of \cite[Janes et al.
(1988)]{Janes_etal88} and \cite[Janes \& Phelps
(1994)]{JanesPhelps94}.
\begin{figure}
% \vspace*{-2.0 cm}
\begin{center}
 \includegraphics[width=3.4in]{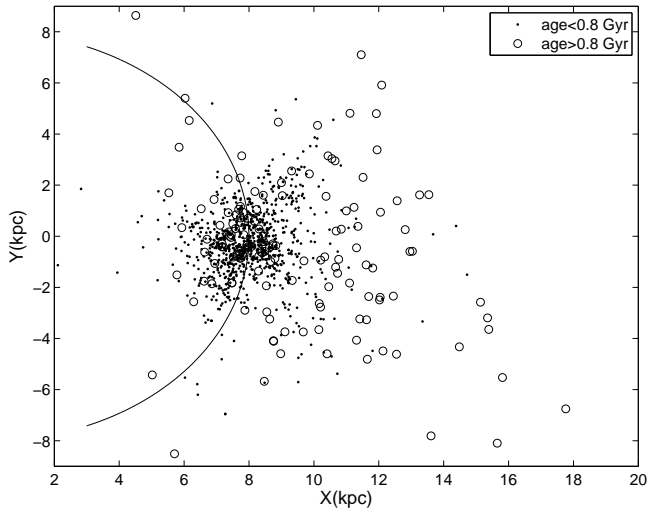}
% \vspace*{-1.0 cm}
 \caption{Spatial distribution of OCs in the Galactic $(X, Y)$ plane. The
open circles and dots are the older and younger clusters,
respectively. The large semi-circle has a radius of 8.0 kpc,
centered on the Galactic center.}
   \label{fig1}
\end{center}
\end{figure}

In our kinematic sample, we have compiled a total of 369 OCs for
which distance and both proper motion and radial velocity data are
available. Here, in left-hand panel of Fig. 2, we plot their
projected velocity onto the disk, whilst the right-hand panel
shows the averaged rotation velocity of OCs as a function of the
galactocentric distance.
\begin{figure}
% \vspace*{-2.0 cm}
\begin{center}
 \includegraphics[width=3.4in]{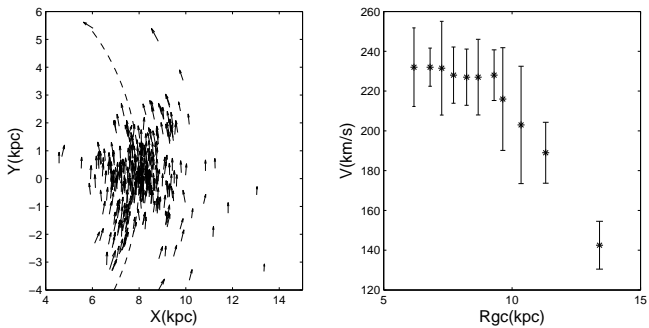}
% \vspace*{-1.0 cm}
 \caption{{\it Left:} Velocity projection on the Galactic plane of 369 clusters. The
Sun is located at $(X=8, Y=0)$ kpc. {\it Right:} Rotation velocity
of these OCs as a function of galactocentric distance}
   \label{fig2}
\end{center}
\end{figure}

Based on our kinematic dataset we can also constrain the Galactic
differential rotation, as well as the Galactic radial motion
parameters at the solar neighborhood. We have compiled a subsample
of 117 clusters with heliocentric distances of $0.5-2.0$ kpc and
ages younger than 0.8 Gyr, which can be considered as typical
thin-disk objects in the Galaxy, and for which the Oort theory is
applicable. The kinematical parameters determined from these
clusters can be used to represent the kinematic properties of
thin-disk objects in the vicinity of the Sun. Thus, we deduced the
Galactic components of (i) the mean heliocentric velocity of the OC
system, $(u1,u2,u3) = (-16.1 \pm 1.0, -7.9 \pm 0.4, -10.4 \pm 1.5)$
km s$^{-1}$, (ii) the characteristic velocity dispersions,
$(\sigma_{1},\sigma_{2},\sigma_{3}) = (17.0 \pm 0.7, 12.2 \pm 0.9,
8.0 \pm 1.3)$ km s$^{-1}$, (iii) the Oort constants, $(A, B) = (14.8
\pm 1.0, -13.0 \pm 2.7)$ km s$^{-1}$ kpc$^{-1}$, and (iv) the radial
motion parameters, $(C,D) = (1.5 \pm 0.7, -1.2 \pm 1.5)$ km s$^{-1}$
kpc$^{-1}$ (\cite[Zhao et al. 2006]{Zhao_etal06}). The parameters
determined from these clusters have accuracies significantly greater
than those obtained from other groups of clusters.

\section{The disk abundance gradient based on open clusters}

OCs can also be used as a powerful tool to understand whether and
how the spatial abundance distribution changes with time, because
OCs have formed at all epochs and since their ages, distances, and
metallicities can be derived more reliably than the equivalent
parameters of the field stars.

We have compiled the full OC sample, containing 144 objects, with
metallicity, distance and age parameters. From this sample, we
obtain a radial metallicity gradient of -0.058 $\pm$ 0.006 dex
kpc${^-1}$ (Chen et al., in prep.), for galactocentric distances
ranging from about 7 kpc to 17 kpc. By dividing the clusters into
young and old subsamples (see also \cite[Chen et al.
2003]{Chen_etal03}), we find that the corresponding gradients are
significantly different, as shown in the upper panel of Fig. 3. That
is, the gradient is steeper in the past, and shallower for younger
clusters. In the bottom panel of Fig. 3, the gradients of the inner
and outer subsamples have similar values.

\begin{figure}
% \vspace*{-2.0 cm}
\begin{center}
 \includegraphics[width=3.4in]{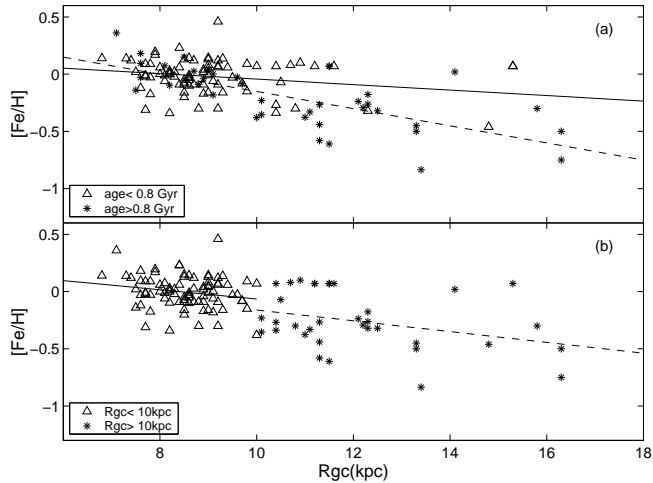}
%\includegraphics[]{7105_chen_3.eps}
% \vspace*{-1.0 cm}
 \caption{{\it (a)} Time evolution of the [Fe/H] gradient. Triangles and stars represent younger and older OCs, respectively. {\it (b)} Gradients for
 inner-disk and out disk clusters. (\cite[Chen et al. 2003]{Chen_etal03})}
\label{fig3}
\end{center}
\end{figure}

This abundance gradient result is consistent with those from HII
regions (\cite[Deharveng et al. 2000]{Deharveng_etal00}) or
planetary nebula data (\cite[Maciel et al. 2006]{Maciel_etal06}).

In a recent review, \cite[Maciel et al. (2006)]{Maciel_etal06}
combined abundance data from planetary nebula and OC samples to
provide observational constraints on disk chemical evolution models.
As shown in fig.10 of their paper, these authors investigated the
abundance gradient as a function of time by combining the results
derived from OC data by \cite[Friel et al. (2002)]{Friel_etal02},
and by \cite[Chen et al. (2003)]{Chen_etal03}. The theoretical
tracks of \cite[Chappini et al. (2001)]{Chiappini_etal01} and
\cite[Hou et al. (2000)]{Hou_etal00}, respectively, are also shown
in that figure. The differences between these two model predictions
are rather large. In Hou et al. (2000)'s model, based on an
exponentially decreasing infall rate and an "inside-out" scenario
for the formation of the Galactic disk, a rapid increase of the
abundance at early times in the inner disk caused a steep gradient.
As the star formation activity migrates to the outer disk, the
abundances are enhanced in that region, so that the gradients
flatten out. In the model used by Chiappini et al. (2001), two
infall episodes are assumed to form the halo and the disk. The disk
is also formed in an "inside-out" formation scenario, in which the
time-scale is a linear function of the galactocentric distance. In
general, some steepening of the gradient is predicted by that model.
It seems that Hou et al. (2000)'s model is better supported by the
observational data, however.

Thus, based on OC data we may constrain the disk chemical
evolution such that in the early stage of disk formation it showed
a steeper abundance gradient, while later on this gradient became
flatten out. Anyway, these inferences are not very conclusive,
since we still do not have a sufficiently large old and outer disk
cluster sample. The compilation of such a sample is critical for
these statements to be made firmer, and we propose to focus on
obtaining such a sample as a priority in this field.

As regards the overall gradient along the disk, \cite[Twarog et al.
(1997)]{Twarog_etal97} proposed a step function for the disk
abundance. That is, inside 10 kpc, there is a shallow gradient,
while beyond 10 kpc the sample is too small to determine if a
gradient exists. They used a sample of 14 OCs in the outer disk.
However, given that we now have 34 OCs beyond 10 kpc, from our data
a continuous linear gradient is favoured instead.

\section{Very young open clusters and the effects of mass segregation}
In some previous work on young open clusters, including the Orion
Nebula Cluster (\cite[{Hillenbrand 1997}]{Hillenbrand97}) and R136
(e.g. \cite[Cambell et al.1992]{Cambell92}, \cite[Brandl et al.
1996]{Brandl_etal96}) the authors found evidences for mass
segregation. However, the question is, for such young star clusters,
whether these effects are mainly a dynamical result or possibly
primordial in nature.

Recently, we investigated two very young open clusters, NGC 2244 and
6530, both with ages younger than 5 Myr. As an example, here we show
some results for NGC 6530 (\cite[Chen et al. 2007]{Chen_etal07}).

The observational data were taken from our historical photographic
plates at Shanghai Observatory, with a time baseline of 87 years.
The old photographic plates were taken by the 40-cm refractor at
Sheshan observing station.

From this photographic data, we determined the proper motion for 495
stars in a one square degree region centered on the cluster, and
estimated their membership probabilities. Since these cluster stars
are located at the same distance, and because most of them are main
sequence stars, we can use luminosity instead of mass to study the
(mass/luminosity) segregation effects. Fig. 4 shows the cumulative
radial number density profile for bright and faint cluster stars,
which shows that brighter (or massive) stars are more concentrated
towards the inner part of the cluster. this indicates an evident
luminosity or mass segregation effect. A similar result was obtained
for our other sample cluster, NGC 2244 (\cite[Chen et al.
2007]{Chen_etal07}).

\begin{figure}
% \vspace*{-2.0 cm}
\begin{center}
 \includegraphics[width=3.4in]{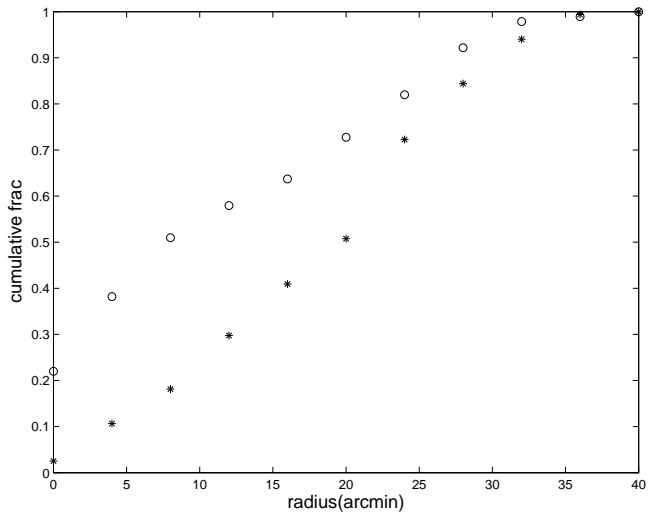}
%\includegraphics[]{7105_chen_4.eps}
% \vspace*{-1.0 cm}
 \caption{Normalized cumulative
radial number density profile for NGC 6530 members with $m_B \leq
12$ (o) and $m_B > 12$ ($\ast$). Result from Chen et al. (2007)}
\label{fig4}
\end{center}
\end{figure}

We conclude that both OCs, NGC 6530 and NGC 2244, show clear
evidence of mass (luminosity) segregation. These two clusters were
too young for this to have only been caused by dynamical relaxation
processes, and the observed segregation might be due to a
combination of both initial conditions and dynamical evolutions.
Therefore, our results supporting the ``primordial mass
segregation'' scenarios, such as the ``competitive accretion''
model. (see \cite[Larson 1991]{Larson91}, \cite[Bonnell et al.
2001a]{Bonnell_etal01a}, \cite[Bonnell et al.
2001b]{Bonnell_etal01b})

\section{The LOCS -- LAMOST Open Cluster Survey project}

Regarding the present status of our OC database, we need a much
larger sample with both abundance and three-dimensional spatial
motion data in order to study the global disk dynamical and
chemical evolution. In particular, for our study of the disk
chemical evolution, we need more outer-disk cluster data so that
we can determine the abundance gradient evolution decisively.
Equally important, a unified observational data set for different
clusters is very much needed. As for a single cluster, in many
cases only a few stars are observed spectroscopically to derive
the average abundance or radial velocity values. On the whole, a
large and homogenous spectroscopic data set for OCs is as yet
non-existent.

To improve the situation, there are some important OC projects
ongoing or in the planning phase. These includes, for example, WOCS
(The WIYN Open Cluster Study; \cite[Mathieu 2000]{Mathieu00}), which
aims to obtain comprehensive and definitive astrometric, photometric
and spectroscopic databases for tens of fundamental clusters; SOCS
(SEGUE Open Cluster Survey), which -- as part of the AS-2 project --
plans to obtain radial velocities and metallicities for more than
100 OCs. The Italian BOCCE project (\cite[Bragaglia
2006]{Bragaglia_etal06}), taking photometric and high-resolution
spectroscopic observations of 30 clusters, will embark on a detailed
investigation of their ages, distance and abundance properties.

Recently we have proposed the LOCS project -- the LAMOST Open
Cluster Survey.

The Large Sky Area Multi-Object Fiber Spectroscopic Telescope
(LAMOST) project (see http://www.lamost.org/en/) is one of the
National Major Scientific Projects undertaken by the Chinese Academy
of Science. LAMOST is a 4-meter aperture quasi-meridian reflecting
Schmidt telescope, with a large field of view of about 20 square
degree, and 4000 fibers.

The telescope will be located at the Xinglong Observing Station of
the National Astronomical Observatories. At present, a small system
with an effective 2-meter aperture instrument has already obtained
first light. The full system is scheduled to finish its final
assembly in early 2009.

For the LOCS, the main advantage of using LAMOST is that it can be
most efficient for an OC survey. With its multiple fibers and large
field of view, we can complete spectroscopic survey observation for
at least one cluster per observing night. This type of survey
observations will have deep enough magnitude limits to reach most of
the old clusters within $5-8$ kpc.

We expect the LOCS, in about 4-years of observations, to survey a
total of around six hundred OCs, each covering one square degree of
sky area. We wish to obtain a homogeneous metallicity and radial
velocity sample of stars in each cluster region. This will lead to
reliable membership determinations, which will -- in turn -- form a
fundamental basis for OC studies. Obtaining many more abundance and
kinematic properties of old and distant OCs provides crucial
constraints for modelling of the structure and evolution of the
Galaxy.

From the selected OC targets, we choose 150 OCs with suitable ages
and distances as the first-priority sample for LOCS observations.
Meanwhile, thirteen fully studied OCs have been chosen as standard
objects for a comparison purpose. We have already prepared the input
catalog for stars in the selected cluster sky area, and a
preliminary but systematic membership estimation for candidate clusters is ongoing.\\

LC, JLH and JLZ are supported by the NSFC (Grant Nos. 10773020,
10573028,10333050, 10333020,10573022) and NKBRSFG 2007CB815403 and
403. RdG was partially supported by an ``International Joint
Project'' grant, jointly funded by the Royal Society in the UK and
the Chinese National Science Foundation (NSFC). This research has
made use ofthe WEBDA database, operated at the Institute for
Astronomy of the University of Vienna.

\end{document}